# AI and Citizen Science for Serendipity

Marisa Ponti[1] , Anastasia Skarpeti[2] , Bruno Kestemont[3]

[1] Department of Applied Information Technology, University of Gothenburg, Sweden
[2] Norwegian University of Science and Technology (NTNU), Ålesund, Norway
[3] Foldit Player, Go Science Team

**Abstract.** It has been argued that introducing AI to creative practices destroys spontaneity, intuition and serendipity. However, the design of systems that leverage complex interactions between citizen scientists (members of the public engaged in research tasks) and computational AI methods have the potential to facilitate creative exploration and chance encounters. Drawing from theories and literature about serendipity and computation, this article points to three interrelated aspects that support the emergence of serendipity in hybrid citizen science systems: the task environment; the characteristics of citizen scientists; and anomalies and errors.

**Keywords:** Artificial Intelligence, Citizen Science, Computational Creativity, Machine Learning, Serendipity, Systems Design

## 1    Introduction

Polyak (2020) argued that introducing AI to creative practices destroys spontaneity, intuition and serendipity in favour of deliberate and premeditated outcomes. However, the design of systems that leverage complex interactions between citizen scientists and computational AI methods have the potential to facilitate creative exploration and chance encounters.

Citizen science – when the general public is actively engaged in research tasks – is already well established in fields such as astronomy and astrophysics, ecology and biodiversity, archaeology, biology, and neuroimaging (Vohland et al., 2021). Citizen scientists are often untrained amateurs recruited by scientists to collect or classify large volumes of data, or solve challenging puzzles (Gura, 2013). Citizen scientists significantly outnumber professional scientists and are rarely experts in the field of the citizen science projects in which they participate. However, curious and dedicated citizen scientists have often shown abilities that can lead to serendipitous encounters (Beaumont et al., 2014). The sheer amount of data potentially generated by citizen science projects, in combination with large numbers of participants, can result in serendipitous discoveries, such as the formation of a novice star, the discovery of a new species, the detection of the human neurons, or the formation of a new protein (Parrish et al. 2019). For example, citizen scientists' serendipitous discoveries led to the most important findings of Galaxy Zoo, an astronomy project focused on the morphological classification of galaxies (Marshall et al. 2014). This includes the discovery of the ''Green Pea'' galaxies by participants who noticed significant green blobs when classifying images in a million-galaxy dataset (Trouille, Lintott, & Fortson, 2019).

So far, few studies have investigated how computational methods such as machine learning (ML) could facilitate serendipity in citizen science as well as optimizing accuracy and efficiency (Trouille, Lintott, & Fortson, 2019).

This paper considers the design of hybrid citizen science systems to demonstrate their potential for serendipitous discovery. It first clarifies the meaning of serendipity then sets out three main aspects relevant to the emergence of serendipity in hybrid systems in citizen science.

## 2    Serendipity and design

Serendipity can be defined as "... making discoveries, by accidents and sagacity, of things [one is] not in quest of ..." (1754, quoted in Merton & Barber, 2004, p. 2). Serendipity can take different forms and occur in different ways but is consistently associated with *unexpected* and *positive* personal, scholarly, scientific, organizational, and societal events and discoveries (McCay-Peet & Toms, 2015).

Several cognitive models have been developed describing the process of serendipitous discovery (e.g., Cunha, 2005; McCay-Peet & Toms, 2015). These models differ in terms of where in the process an "insight" takes place but all focus on an insightful, cognitive connection made by a 'serendipitous' individual. However, Copeland (2019) criticized these cognitive perspectives because they fail to consider the environment in which individual discoverers operate.

Yaqub's (2018) taxonomy proposes four types of serendipity - theory-led, observer-led, error-borne, and network-emergent - which highlight that it is multifaceted and not restricted to basic science but can occur in applied research and technological development. They also suggest that serendipity is not random since observation routines and instrumentation may affect what an individual notices.

Three factors suggest there is an opportunity to design computational systems that facilitate citizen scientists' creative exploration and chance encounters:

1. **Algorithms**, which have long been recognized to support serendipity by surfacing interesting connections; providing mechanisms to enhance the expertise of would-be discoverers and making it easier to recognize connections; and enabling the growth – or sharing – of an idea so it can be developed by those more keenly interested in the connection (André, Teevan, and Dumais, 2009).

2. **Technology design**, which can influence the possibility of serendipity. Reporting on positive experiences with smart services designed for chance encounters alongside exact retrieval, Danzico (2010) posited that designers had simply created opportunities for people to find their own pathways – it was up to them to find chance. Technologies can also support a shift from the reliance on serendipitous individuals to hybrid combinations of curious humans well-suited to serendipitous encounters and technologies.

3. **ML**, which is already used in citizen science projects to process massive amounts of data more accurately and efficiently (Trouille, Lintott, & Fortson, 2019). For example, astronomy discoveries are often driven by serendipitous encounters with outliers but as databases increase in size and complexity, astronomers' ability to manually scan the data and make discoveries decreases – instead, machine-based methods are needed to mine the data to have any hope of identifying anomalous or outlying data (Giles & Walkowicz, 2019).

These three factors are interrelated because technological environments can be designed to allow interactions that intentionally encourage chance encounters. In a culture aiming to design systems to be less error prone and more exact, with specific interactions and precise retrieval, it is important to explore how chance encounters can still be fostered by designing systems that leverage laypeople's problem solving and creativity.

## 3  Facilitating serendipity

Hybrid systems should, then, be designed to stimulate participants' serendipitous encounters, rather than to detect serendipitous occurrences without human intervention. Serendipity should be considered a collective endeavor comprising individual cognitive agency coupled with external representations and cognitive artifacts, such as ML models. The effort should be shared between humans and technology because it is the integration of humans and machines that holds the potential for serendipitous scientific discovery, rather than the former or the latter individually.

Three interrelated aspects are therefore relevant for the development of serendipity in hybrid systems in citizen science: (1) the task environment; (2) the characteristics of citizen scientists, and (3) anomalies and errors. Each of these are examined briefly below.

### 1. Task Environment

The importance of the task environment in which individual agents operate was underscored by Pease et al. (2013), who proposed four characteristics of the discoverers' environments and computational analogs:

  i. *Dynamic world*: Data is presented in stages, not as a complete and consistent whole. This corresponds to streaming from live media such as the web.

  ii. *Multiple contexts*: Information from one context or domain can be used in another. This is a common notion in analogical reasoning.

  iii. *Multiple tasks*: Discoverers are often involved in multiple tasks. This corresponds to threading, or distributed computing, for example.

  iv. *Multiple influences*: All discoveries take place in a social context, and in some examples the "unexpected source" is another person. This corresponds to systems such as agent architectures, in which software agents with different knowledge and goals interact.

These characteristics can be fostered in citizen science projects, which have diverse task environments. For example, in Foldit, a popular online citizen science game in which players are scored on the structure of proteins that they have folded (Cooper, 2014), players are invited to solve new types of puzzles. When playing Foldit, one of the authors acknowledges the use of experience solving problems in other contexts (e.g., in his profession). He also points to the importance of experience playing the game in bringing surprising discoveries. Expert players can be more efficient at solving new types of puzzles because they can "filter out" new sub-problems from ones they have seen before.

The interactions between a great number of individuals in a project also influence the possibility that unexpected observations give rise to new discoveries. For example, Foldit developers improved structural models based on the collective ability of nonprofessional citizen scientists, who made promising observations and important discoveries (Khatib et al., 2019).

## 2. Characteristics of citizen scientists

This is related to the notion of the "prepared mind", which French scientist Louis Pasteur considered the main condition to favor serendipity in the fields of observation (Meyers, 2007). While for Pasteur the prepared mind was connected to prior knowledge on a subject, in citizen science the prepared mind may be seen as openness and curiosity. Furthermore, citizen scientists can come from every walk of life bringing a variety of knowledge, expertise, skills, and talents. All these factors may influence the emergence of serendipity.

Motivation and dedication may also count. Parrish et al. (2019) stressed the importance of time and persistence – the more participants practice, the better they perform. Curious and sagacious amateur experts with "relatively rare skills and knowledge" are also well-situated to make serendipitous discoveries (Sauermann and Franzoni, 2015, p. 1).

## 3. Anomalies and errors

Anomalies and errors have always played prominent roles in accounts of serendipity (Yakub, 2018; Merton, 1948). As the above discussion of Galaxy Zoo shows, the scale of participation possible in citizen science projects can help to discover anomalies even in large datasets. Elsewhere in a marine citizen science project, for example, unusual clips are brought to the attention of participants for classification (Koster Seafloor Observatory, n.d).

Surprises are also important for human engagement. People may be less exposed to chance or less inclined to try new things if their tasks are planned and designed in such a way that there are no discoveries or surprises (Danzico, 2010). In Foldit, for example, some pioneer players argued in defense of trials and errors as the excessive reliance on "helper algorithms" would reduce creativity, resulting in deterministic solutions and limiting discovery of new protein types ("What's the point? Can't this just be automated using recipes?", n.d.).

Scoring mechanisms can also inadvertently penalize unexpected events. In Foldit, "expected" solutions tend to generate higher scores meaning that players use their knowledge and intuition to persevere on low scored but promising "unexpected" solutions, turning them into higher score "expected" results.

# 4    Conclusion

This article presents three aspects to be taken into account to design serendipity-oriented hybrid systems. This is important because the integration of humans and machines in citizen science has been acknowledged to produce results superior to either one alone, while allowing for serendipitous discovery (McClure et al., 2020; Beaumont et al., 2014). Algorithms, technology design and ML present opportunities for serendipity, if applied with care. Future citizen science projects should therefore reflect on the technology environment, characteristics of citizen scientists and anomalies and errors to encourage serendipity. Further research could include a survey of serendipity-oriented algorithms to examine existing approaches used in science and innovation; deeper investigation of delegation of tasks in human-machine integration; and recommendations for the design of computational assistance to support serendipity and human agency.

## Acknowledgements

This work was supported by Marianne and Marcus Wallenberg Foundation, grant no. 2018-0036.